\documentclass[twocolumn,11pt]{style-atc-arxiv} 

\newcommand{\spaceexpander}{1.2} 	

\usepackage{listings}

\newcommand{\thisCategory}{Editorial} 

\newcommand{\thisYear}{2021}
\newcommand{\thisVolume}{1}
\newcommand{\thisIssue}{1}
\newcommand{\thisFirstPage}{1}

\newcommand{\Nine}{Nine\;}
\newcommand{\nine}{nine\;}

\title{\Nine Challenges in Modern Algorithmic Trading and Controls} 
\runningtitle{Nine Challenges of ATC} 	

\author{Jackie Shen}	

\cauthor{Jackie Shen, ojs-atc@deepquantech.com}

\affiliation{Deep QuanTech, LLC, New York, NY 10025, USA}

\abstract{This editorial article partially informs the algorithmic trading  community about launching of the new journal \href{https://atc.deepquantech.com/}{\textit{Algorithmic Trading and Controls}} (ATC). ATC is an online open-access journal that publishes novel works on algorithmic trading and its control methodologies. In this inaugural article, we discuss nine major challenges that contemporary Algo trading faces. There is nothing superstitiously magical about the number ``nine," but so is any other one. Several of these challenges are at the strategy level, including for example, trading of illiquid securities or optimal portfolio execution. Others are more at the level of risk management and controls, such as on how to develop automated controls, testing and simulations. The editorial views could be inevitably personal and biased, but have been explored with the most innocent intention of contributing to this important field in modern financial services and technologies. 
}

\keywords{Algos, liquidity, portfolio, correlation, special days, derivative pricing, universe, clustering, machine learning, auctions, shortfall, transaction cost, unit test, regression test, simulation, automated controls} 

\begin{document}

\maketitle 				
\makeaffils				

\renewcommand{\baselinestretch}{\spaceexpander} 


\section{Introduction} 

\newcommand{\wwwATC}{https://atc.deepquantech.com}
\newcommand{\wwwThreeLines}{https://wiki.treasurers.org/wiki/Three_Lines_of_Defence_Model}
\newcommand{\wwwFED}{https://www.federalreserve.gov}
\newcommand{\wwwPRA}{https://www.bankofengland.co.uk/prudential-regulation}

\newcommand{\wwwRFQ}{https://en.wikipedia.org/wiki/Request_for_quote}

By this inaugural article, the new journal \href{\wwwATC}{\textit{Algorithmic Trading and Controls}} (ATC) is formally rolled out. 

ATC is a \textit{discretionary, moderately for-profit}, and \textit{online open-access} journal that publishes novel works on algorithmic trading (\textit{a.k.a.} ``Algo" or ``Algo trading") and related control frameworks. Automated Algo trading generates both revenues and risks, and hence the importance of automated controls should never be underestimated. We refer interested readers to the website of ATC for more details: \href{\wwwATC}{https://atc.deepquantech.com}.

In this editorial article, we discuss nine major challenges that contemporary Algo trading faces. Before diving into the details, we first clarify that as for the journal ATC, we restrict the notion of ``Algo trading" to two most common and influential activites: optimal trade execution and market making. The latter covers both exchange market making and automated over-the-counter (OTC) liquidity provision such as \href{\wwwRFQ}{Request-For-Quote} (RFQ). Both the journal ATC and this editorial article do not intend to cover \textit{purely opportunistic} trading activities that seek alphas or arbitraging opportunities for a principal account. The signals and strategies involved in such trading activities are \textit{confidential} and \textit{proprietary}, and by default prohibited from publishing. 

In terms of asset classes, we focus more on well-regulated markets such as of equities, futures, rates products, or liquid foreign exchanges (FX). Whenever applicable, we also comment on others such as OTC derivatives or credits.  This article also focuses primarily on the markets in North America, esp. in the United States. 
 
The \nine challenges being explored are as follows.
\begin{enumerate}[label={[}\arabic*{]}]
\item Trading Illiquid Securities

\item Optimal Portfolio Execution 

\item Clustering of the Tading Universe

\item Handling of Special Days

\item Real-Time Pricing of Derivatives

\item Trading in (Close) Auctions

\item Transaction Cost Models

\item Automated Controls for Automated Trading

\item Full-Scale Testing and Simulation
\end{enumerate}
Some of these challenges call for either strategy revamping or more intelligent data analytics, while others are more concerned with risk controls or robust testing and simulation frameworks. Both are in scope for this new exciting journal and require broker-dealers or trading houses to make further investments in talents, analytics, infrastructures, and the \href{\wwwThreeLines}{Three Lines of Defense}. 

Finally, an editorial article of such nature could be inevitably personal and biased. But its purpose is to initiate further healthy dialogues within the Algo trading community, which includes fund managers, broker-dealers, trading agents or houses, and regional supervisory functions such as the \href{\wwwFED}{Federal Reserve Board} or the \href{\wwwPRA}{Prudential Regulation Authority}.  

Algo trading has been playing an increasingly vital role in the modern landscape of financial technologies and services, and profoundly impacting both the main street and Wall Street. As a result, it can no longer be operated in the conventional style of black boxes, and must start to promote a healthy culture of open discussions, information sharing, and collaborations.


\section{The \Nine Challenges for ATC}

We now elaborate on the nine challenges. The order is somewhat at will due to the relative independence.

\newcommand{\wwwADV}{https://www.investopedia.com/terms/a/averagedailytradingvolume.asp}
\newcommand{\wwwBOSpreads}{https://en.wikipedia.org/wiki/Bid-ask_spread}
\newcommand{\wwwVWAP}{https://en.wikipedia.org/wiki/Volume-weighted_average_price}
\newcommand{\wwwTWAP}{https://en.wikipedia.org/wiki/Time-weighted_average_price}
\newcommand{\wwwPegging}{https://en.wikipedia.org/wiki/Order_(exchange)\#Peg_orders} 
\newcommand{\wwwBestEx}{https://en.wikipedia.org/wiki/Best_execution}
\newcommand{\wwwBestExMifid}{https://www.esma.europa.eu/document/best-execution-under-mifid}
\newcommand{\wwwIS}{https://en.wikipedia.org/wiki/Implementation_shortfall}
\newcommand{\wwwNBBO}{https://www.investopedia.com/terms/n/nbbo.asp}
\newcommand{\wwwESMA}{https://www.esma.europa.eu/}
\newcommand{\wwwMuniBond}{https://en.wikipedia.org/wiki/Municipal_bond}

\subsection{Trading Illiquid Securities}
\label{sub:1-illiquid}

On the market side, the limit order books of an illiquid name are typically ``thin" and active market participants are also very limited, e.g., mostly registered market makers who are obliged to post two-sided quotes. Such poor liquidity qualities make these names highly sensitive to supply-demand imbalances.  

As a result, most illiquid names invariably share these macro characteristics: lower average daily volumes (\href{\wwwADV}{ADV}), wider average \href{\wwwBOSpreads}{bid-offer spreads}, and higher volatilities. All these negative factors aggravate \href{\wwwIS}{implementation shortfalls} (IS) whenever trading these names. 

On the trading side, the constraint of completion (within a given execution horizon) raises a major hurdle for Algos. Any passive waiting and inaction ``now" may pile up positions for ``later" stages and hence increase the \textit{squeezing} risk and impact costs.  For this reason, traders may better turn to Algos with flat execution profiles such as \href{\wwwVWAP}{VWAP}/\href{\wwwTWAP}{TWAP} or Participation of Volumes (PoV). Further diligence must be exercised because the intraday volume profiles of illiquid names demand either longer time bins to accumulate sufficient volumes, or otherwise to be interpreted in probabilistic manners.

For equities, dark pools (wherever available in the regional markets) offer alternative venues to minimize information leakage or seek extra liquidities. They are indeed indispensable for many Algos crafted specifically for illiquid or small-cap names.
When placing \href{\wwwPegging}{pegging orders} in dark pools (e.g., pegged to the mid of the \href{\wwwNBBO}{National Best Bid and Offer} (NBBO)), Algo designers must pay extra attention to monitor the dynamics and toxicity of the pegged prices. In the scenario of a single market maker, for example, the prices may merely reflect the inventory pressure being experienced by the market maker at the moment, and may not necessarily reflect the ``true" values (TV).

For some less liquid fixed-income products, execution usually hybrids agency and principal trading. An execution Algo constantly monitors external venues as well as the estimated TVs and spreads of the agent herself. Whenever the agent can offer prices improved from external venues, the Algo may execute some portions in the \textit{principal capacity} to benefit a client. Such practice must be governed by the \href{\wwwBestEx}{Principle of Best Executions} universally required by regulators (e.g., \href{\wwwBestExMifid}{MiFID} of the European Securities and Markets Authority (\href{\wwwESMA}{ESMA})). This demands robust market data connections with low latency, as well as accurate real-time computation of the TVs and spreads, etc. 

To automate and integrate all the aforementioned logic and signals remains a major challenge for Algo developers, in order to most effectively trade small-cap names or many illiquid fixed-income products. For instance, to automate the trading of \href{\wwwMuniBond}{municipal bonds}, many of which are illiquid, the main hurdle turns out to be very elementary - how to properly estimate their TVs in real time when actual trades are very sparse and hence credible footholds for pricing are simply not there.

\subsection{Optimal Portfolio Execution}

\newcommand{\wwwPortfolioTheory}{https://en.wikipedia.org/wiki/Modern_portfolio_theory}
\newcommand{\wwwEMA}{https://en.wikipedia.org/wiki/Moving_average\#Exponential_moving_average}
\newcommand{\wwwETF}{https://en.wikipedia.org/wiki/Exchange-traded_fund}
\newcommand{\wwwFICC}{https://wiki.treasurers.org/wiki/Fixed_Income_Clearing_Corporation}
\newcommand{\wwwDynProg}{https://en.wikipedia.org/wiki/Dynamic_programming}

Unstructured portfolios can be executed using \textit{asynchronous} single-name Algos. By ``unstructured," we informally refer to baskets that have been formed \textit{without} any systematic objectives in mind, e.g., delta neutrality or long-short balance. 

For a sizable portfolio whose execution spans over a sufficiently long time window, execution risk can be reduced if the individual names happen to hedge among themselves to some degree. This is usually the case for a {\em structured} portfolio, for instance, one resulted from index rebalancing when some new names are to be acquired and an equal amount of selected old names to be liquidated.   

In the \href{\wwwPortfolioTheory}{modern portfolio theory}, hedging is quantified by the correlations and volatilities of the names. They can be calibrated directly from the historically observed returns, or indirectly assembled from  multi-factor models such as the Barra\textsuperscript{\texttrademark} or Axioma\textsuperscript{\texttrademark}. 

Conventionally these are often end-of-day (EOD) models. One major challenge is how to revamp the EOD risk models to land on effective \textit{real-time} risk models. A popular practice is to use the EOD correlation models as the backbone, assuming that correlations vary slowly over longer horizons.  The con of this assumption is that EOD correlations are typically calibrated over longer horizons and may miss any emerging correlating patterns for the purpose of \textit{intraday trading}. Such scenarios can emerge when a given name starts to break away from a cohort due to some major breakthroughs of products or services, e.g., a pharmaceutical company with an important new drug approved, or a public company announced to be included in a popular market index. The information has been released, but classical EOD models may act too slow to reflect it in a timely manner. Using weighting schemes like the \href{\wwwEMA}{exponential moving average} (EWA)) helps catch up in reaction but still acts too passively for the purpose of intraday trading.

Calibrating intraday volatilities imposes another challenge. For any given single name, {\em static profiling} is to establish a static volatility curve $\sigma(t)$ so that, for example, $\sigma(10:18\mathrm{am})$ denotes the average volatility at 10:18am. Such profiles can be prepared overnight and stand by ready before a day-trading session commences. It is a stable and predictable tool, but may lose touch with the intraday reality of a particular given day. A more ideal solution would turn to {\em dynamic profiling} when the entire intraday curve is not pre-calculated but gradually rolled out. At each ``current" time $t$, the future profile $\sigma(t:EOD)$ can be modeled as a stochastic process or updated belief based on what has been observed in the market ``so far."  This can be computationally more expensive but surf well with real-time market waves. 

Away from risk correlations, correlations among {\em impact costs} have been largely muted in both academic and industrial works. The assumption of {\em independent} impact costs might approximate well for most individual names. But there are scenarios well worth further data-driven studies.  For example, suppose that a portfolio contains both a single common stock named SABC and an \href{\wwwETF}{exchange-traded fund} (ETF) named FXYZ that has SABC as one of its {\em sizable} constituents (e.g., SABC=Exxon Mobil Corp. and FXYZ=XLE - Energe Select SPDR Fund). Thanks to index arbitragers,  any sudden push-ups of SABC can be transferred to FXYZ almost instantaneously, and vice versa. As a result, the impact costs of trading sizable amounts of SABC and FXYZ could be intricately entangled, potentially leading to non-negligible and verifiable observables. 

To organically integrate all these risk and cost analytics into coherent portfolio execution models imposes another major challenge. Such Algo models must be mathematically tractable and computational feasible and efficient. Among them, self-contained \href{\wwwDynProg}{dynamic programming models} are notoriously harder.

Finally, as broker-dealers and trading agents become more enthusiastic in integrating and unifying their platforms and offering \textit{cross-asset} trading Algos, it is another challenge to optimally trade portfolios containing multiple but correlated asset classes, e.g., common stocks, ETFs, futures, options, or general \href{\wwwFICC}{FICC} products.

\subsection{Clustering/Classification of the Universe}

\newcommand{\wwwAltData}{https://en.wikipedia.org/wiki/Alternative_data_(finance)}
\newcommand{\wwwValueGrowth}{https://www.finiki.org/wiki/Value_vs_growth}
\newcommand{\wwwML}{https://en.wikipedia.org/wiki/Machine_learning}
\newcommand{\wwwStyleFundamental}{https://en.wikipedia.org/wiki/Fundamental_analysis}
\newcommand{\wwwStyleTechnical}{https://en.wikipedia.org/wiki/Technical_analysis}
\newcommand{\wwwImputation}{https://en.wikipedia.org/wiki/Imputation_(statistics)}

The security universe covered by a typical investment bank, execution agent, or registered market maker is usually vast, potentially including tens of thousands of different names depending on asset classes. This is especially true for global equities trading.

Clustering helps organize trading universes and drastically reduces operating complexities. In a stationary trading environment, clustering can substantially improve operational efficiency by sharing a set of common strategies, parameter factories, implementations, or risk controls within individual clusters. In an emergency scenario, clustering can also offer a default framework or procedure for automated handling, e.g., instantaneous \href{\wwwImputation}{imputation} of certain risk characteristics when data servers or connections are experiencing unauthorized naps or unexpected glitches. 

Conventional frameworks or risk models have already been able to offer some rudimentary schemes of clustering or classification, e.g., via the directions of sectors, industrial groups, or fundamental risk drivers such as market capital sizes, \href{\wwwValueGrowth}{value vs. growth}, etc. 

Modern \href{\wwwML}{machine learning} (ML) techniques  probably can offer more. For the purpose of intraday optimal execution or continuous market making, an overnight process of clustering and classification is more ideal than the traditionally \textit{static} segmentation schemes, such as those merely driven by sectors or industrial groups. Here the main challenges are to sort out all security characteristics, be them \href{\wwwStyleFundamental}{fundamental} or \href{\wwwStyleTechnical}{technical}, that are most relevant to intraday trading for either optimal execution or market making. ML-driven clustering may also have to accommodate conventional risk models and to set proper clustering objectives. Unlike prevailing risk models, non-numerical  characteristics can also be accommodated using modern ML techniques, such as categorical feature variables or those derived from \href{\wwwAltData}{alternative data} like investment sentiments on social media.

\subsection{Handling of Special Days}

\newcommand{\wwwIndexReb}{https://en.wikipedia.org/wiki/Index_fund}
\newcommand{\wwwTripleWitching}{https://www.investopedia.com/terms/t/triplewitchinghour.asp}
\newcommand{\wwwIPO}{https://en.wikipedia.org/wiki/Initial_public_offering}
\newcommand{\wwwFuturesRoll}{https://www.cmegroup.com/trading/equity-index/rolldates.html}

There does not seem to exist a formal theory about \textit{special days}, but industrial practitioners know that they ought to be treated differently. In the modern era when technologies can facilitate quicker and more adaptive responses to market dynamics, indeed they deserve more customized and responsive strategies.  

Well-known examples include the Fed Announcement Days, Half Trading Days, Month-End or Quarter-End Days, \href{\wwwIndexReb}{Index Rebalancing} Days, \href{\wwwTripleWitching}{Triple Witching} Days, etc. Trading patterns are more pronounced on certain special days than on some others. Also days like Index Rebalancing may observe more salient impacts on certain \textit{specific} individual names, i.e., the new joiners and dropouts for an index.

On special days, trading patterns may differ in all sessions, including for instance, the continuous core session, and  open and close auctions. As a result, the corresponding trading parameters should be prepared using proper statistical methods or machine learning techniques. During real-time implementation of a special day, trade beliefs and forecasting must be further updated based on the specially calibrated models, parameter factories, and the actually observed market dynamics. Each special day may assume its very  special identities.

An equally stimulating notion, though less popularly implemented in the Algo trading industry, is ``Special Periods."
Special periods may be anchored around special days, for example, the first trading week or month of an \href{\wwwIPO}{Initial Public Offering} (IPO) for equities, or the week near a \href{\wwwFuturesRoll}{roll date of a major index futures} for futures, etc. They may signify the periods of \textit{known} transitions and uncertainties, and hence are more tractable than latent periods in general regime-switching models. 

Special days or periods represent the non-stationary moments of the life cycles of securities or trading environments, and may present good opportunities for those who master them. Trading with specialized and effective strategies is certainly not a trivial task, and requires special investments in talents and analytics.

\subsection{Real-Time Pricing of Derivatives}

\newcommand{\wwwTaylorExp}{https://en.wikipedia.org/wiki/Taylor_series}
\newcommand{\wwwSplines}{https://en.wikipedia.org/wiki/Spline_(mathematics)}
\newcommand{\wwwBlackScholes}{https://en.wikipedia.org/wiki/Black-Scholes_equation}
\newcommand{\wwwMonteCarlo}{https://en.wikipedia.org/wiki/Monte_Carlo_method}

Traditionally in most investment banks, pricing models have been developed for booking trades \textit{manually}. Traders have to add further overhead premiums such as spreads or commission fees on top of the ``true" prices projected by pricing models. Pricing models and the associated risk analytics are also utilized by  mid- and back-offices for monitoring the end-of-day (EOD) positions and aggregated risks.

With increasing demands on derivatives and more efficient exchange or OTC trading, \textit{automated} trading and clearing of derivatives have gradually become a priority for many investment banks. Among all building blocks, pricing models stand out as the core pillars. 

Here the main challenges include: (A) speeding up computation for pricing and associated risks, and (B) revamping pricing infrastructures, including data connections and servers, to facilitate fast and robust real-time price queries. The two are clearly intertwined. 

Conventional EOD reading and construction of pricing curves, e.g., interest rate curves or credit curves, are too sluggish for intraday and real-time trading.  Ideally a pricing engine must be able to query market data (e.g., money markets, bills, notes, bonds, or swaps and swaptions on rates or credits) in \textit{real time}, and then to construct \textit{on the fly} the implied rates or credit curves. This demands rewiring or upgrade of market data subscriptions and connections, as well as computing engines for curve calibration.

Furthermore, the heavy machinery of \href{\wwwBlackScholes}{partial differential equations} (PDE) for option pricing or \href{\wwwMonteCarlo}{Monte-Carlo} (MC) simulations for exotics has to be re-designed, in order to substantially catch up in speed for real-time dynamic environments. \href{\wwwTaylorExp}{Taylor or asymptotic expansion}, \href{\wwwSplines}{spline interpolation or extrapolation}, and more general approximation techniques probably have to be adopted to reduce frequent calls of the heavy artillery. The second- or minute-time windows of computation must be compressed towards the scale of milli- or micro-seconds for Algo trading, at the sacrifice of some accuracy.  

On the control side, both the Tech Risk Management (TRM) and Model Risk Management (MRM) must step up in scrutinizing the soundness and robustness of these novel infrastructures and real-time pricing logic. \textit{Previous approvals on EOD pricing models do not automatically transfer to real-time models!}

Needless to say, automated derivative trading will become the most exciting area for most investment banks or trading firms. It requires serious investments in the best infrastructures, analytics, and above all, IT talents.

\subsection{Trading in (Close) Auctions}

\newcommand{\wwwNMS}{https://en.wikipedia.org/wiki/National_Market_System}
\newcommand{\wwwDirac}{https://en.wikipedia.org/wiki/Dirac_measure}
\newcommand{\wwwAuctionIndPx}{https://www.investopedia.com/terms/i/indicative-match-price.asp}
\newcommand{\wwwNYSEDOrder}{https://www.nyse.com/article/trading/d-order}

There are \textit{open} and \textit{close} auctions in the US \href{\wwwNMS}{national market system} (NMS). \textit{Intraday} auctions also exist in some regions such as Europe. Auctions provide important alternatives for liquidity sourcing and price formation, and play a critical role in modern-day trading. 

Among all, \textit{close} auctions are becoming the most prominent trading sessions across global markets. Perhaps it can be best justified by this simple keyword - ``completion," that universally governs intraday trading activities, be them high-touch or low-touch. 

When traders or portfolio managers are mandated to liquidate or acquire certain positions before any given EOD, the close auction offers the final \textit{substantive} pool of liquidities.  This applies, for instance, to index fund managers who attempt to minimize fund tracking errors on an index rebalancing day, or to traders on a central risk book who attempt to stay compliant with a firm's internal EOD risk limits and allocations.      

Close auction volumes have been on steady rise and no serious liquid-seeking traders can afford to miss them. For developed markets such as in US or Europe, average close auction volumes have already stepped into the double-digit zone (as a percentage of the average daily volume (ADV)). 

Algos offer a variety of options to traders or clients for effectively tapping auction liquidities. Unless explicitly instructed to complete execution \textit{before} the close, in theory any Algo can offer participation in close auctions. Taking VWAP for instance, a natural way is to treat the close auction volume as a \href{\wwwDirac}{Dirac ``point" mass} and then to allocate the auction participation proportionally.   

In addition, there also exist {\em dedicated} auction Algos that are marketed under the name of ``Target Close (TC)." Different broker-dealers or execution agents may design it with their own objectives and customizable options. Each TC Algo attempts to benchmark against the close price while maximally curbing potential price impact or information leakage. This means that some portions may have to be traded in the continuous core session just before a close auction. 

To best serve the interests of trading clients, all these Algos that tap close liquidities must develop forecasting models on auction volumes and prices, as well as their pre-close dynamics.  They must properly digest information such as imbalance and \href{\wwwAuctionIndPx}{indicative prices} that is being continuously disseminated to the public after a certain time before a close auction (e.g., 3:45pm in US). 

Modern data analytics and machine learning methods can probably improve these predictive models. Traditionally only straightforward statistics have been explored. The main challenges here are that each primary exchange has its own auction roll-out procedures and rules, and that some specific operations could perplex modeling efforts, e.g., special orders like \href{\wwwNYSEDOrder}{NYSE's Closing D Orders}. Finally, it is also nontrivial to seamlessly integrate these predictive analytics into a self-contained and objective-driven optimization problem.

\subsection{Transaction Cost Models (TCM)}

\newcommand{\wwwGravity}{https://en.wikipedia.org/wiki/Gravity}
\newcommand{\wwwNewton}{https://en.wikipedia.org/wiki/Newton's_law_of_universal_gravitation}
\newcommand{\wwwEinstein}{https://en.wikipedia.org/wiki/General_relativity}
\newcommand{\wwwKisselRG}{http://www.kissellresearch.com/krg-i-star-market-impact-model}

In the current article, TCM is restricted to \textit{pre-trade} forecasting models for estimating the transaction costs of trading any proposed positions. We shall reserve TCA, \textit{Transaction Cost Analysis}, for any \textit{post-trade} analysis on the costs of actually executed trades.  The costs due to fees and commissions are out of scope, since they are either published or contracted. In addition, transaction costs here mainly refer to the \textit{impact costs}, not the market risk costs associated with innate market fluctuations.  

In an editorial article like this, savvy readers might have been searching for the keyword ``TCM" from the very start. Indeed, TCM is probably the most celebrated metric in Algo Trading, though this does not mean that it has been thoroughly understood. 

In fact, TCM to the Algo community behaves a bit like the concept of ``\href{\wwwGravity}{gravity}" to the society. For thousands of years, human beings have been aware of the existence of gravity and successfully applied it to important social-economic activities such as measuring the weight of grains for taxing purposes or the weight of gold and silver as currencies. The true enlightenment of gravity, however, did not emerge until \href{\wwwNewton}{Newton} and \href{\wwwEinstein}{Einstein} uncovered the laws behind. 

In the earlier years, broker-dealers or trading agents did openly reveal their TCM models either formally or informally. But the trend is that these models sink deeper and deeper underwater, and become \textit{proprietary} and \textit{confidential}. This is especially true for many emerging TCM models for FICC, such as those for bonds or Foreign Exchanges (FX). Freely accessible TCM models are very rare. For instance, only the Kissell Research Group (KRG) still maintains an open and free \href{\wwwKisselRG}{TCM model under the brand name of ``I-Star,"} at the time when this article is published.

In theory, for any given security there should be a single ground-truth TCM model, which should be kept open, transparent, and accessible to any traders or fund managers. The fact that different firms develop their own \textit{confidential} and \textit{proprietary} TCM models perhaps already suggests something disturbing. Or rather, it may have also signalled the very complexity and ambiguity of the notion of TCM.  Even for post-trade TCA when trade data have been completely observed, it is not so straightforward to carve out the \textit{net} impact costs. 

Let us dip into some light details. Most TCM models seek a function form of:
	\[  TCM = \Phi(Q, [T_0, T_1] \mid s ), \]
where $s$ denotes a given security, $Q$ a targeted buy/sell position in $s$, and $[T_0, T_1]$ a designated execution window, e.g., $Q=60,000$ shares, $T_0=10:00$ am, and $T_1=2:00$ pm. In terms of analytics, the security $s$ supplies all the cost and risk parameters, e.g., average spread $\theta_s$, average daily volume $ADV_s$, and average volatility $\sigma_s$. Hence the expanded function form is given by:
\[
	TCM = \Phi(Q, [T_0, T_1], \theta_s, ADV_s, \sigma_s). 
\]
It is a convenient format for pre-trade cost forecasting, as well as for portfolio optimization when trading costs are factored in. But it does not differentiate among the actual Algos. Such a model often implicitly assumes the VWAP or PoV (Participation of Volume) Algo. A more ideal model should indicate such dependency, i.e., 
\[
	TCM = \Phi(Q, [T_0, T_1], \theta_s, ADV_s, \sigma_s \mid Algo). 
\]
For example, the net impact cost is very different for a typical \textit{front-loading} \href{\wwwIS}{Implementation Shortfall} (IS) Algo that is benchmarked against the arrival price, as versus a more \textit{flat-loading} VWAP Algo. Using an Algo-independent TCM model to project IS impact costs is doomed to be inaccurate. 

The reality is that few broker-dealers or trading houses provide Algo-specific TCM models, to our best knowledge. 

\newcommand{\wwwRefBertLo}{https://doi.org/10.1016/S1386-4181(97)00012-8}
\newcommand{\wwwAlmgrenChriss}{https://doi.org/10.21314/JOR.2001.041}
\newcommand{\wwwShenPreTrade}{https://doi.org/10.1093/amrx/abu007}

\newcommand{\wwwLOB}{https://en.wikipedia.org/wiki/Order_book_(trading)}
\newcommand{\wwwRelativityTheory}{https://en.wikipedia.org/wiki/Theory_of_relativity}
\newcommand{\wwwRSquared}{https://en.wikipedia.org/wiki/Coefficient_of_determination}
\newcommand{\wwwLimitMarket}{https://dx.doi.org/10.2139/ssrn.3574365}

Furthermore, TCM modelling also faces some theoretical challenges. Consider a \textit{schedualing}-based Algo, be it a non-dynamic VWAP or IS Algo, for which the execution path $\displaystyle q_{t \in [T_0, T_1]}$ is pre-scheduled by a suitable optimization model and satisfies
\[
	\int_{T_0}^{T_1} q_t dt = Q, \quad \mbox{with $q_t$ denoting trading speed.}
\]
It is generally held true that the final netted IS, expressed as the basis-point spread over the arrival price, bears the form of:
\[
	IS = C + Z.
\]	
Here $\displaystyle C= TCM=\mathcal{F}(q_{t \in [T_0, T_1]} \mid s)$ is a \textit{deterministic} functional of the execution path $\displaystyle q_{t \in [T_0, T_1]}$, for the given security $s$, and $Z$ is a zero-mean \textit{random} component resulted from the innate stochastic price fluctuations of the market (e.g., Brownians as in most published works including \href{\wwwRefBertLo}{Bertsimas and Lo}, \href{\wwwAlmgrenChriss}{Almgren and Chriss}, or \href{\wwwShenPreTrade}{Shen}, just to name a few). In particular, one has the convenient interpretation of the TCM: $TCM = C = E[ IS ]$. 

In reality, even for a given \textit{deterministic} schedule $\displaystyle q_{t \in [T_0, T_1]}$, $C$ is still \textit{stochastic}. The cost component $C$ involves the complex interactions of numerous real-time factors, including the dynamics of the \href{\wwwLOB}{limit order books}, the strategy of \href{\wwwLimitMarket}{allocating marketable vs. limit orders}, and the usage of dark or grey venues and different order types. Some of these variables are latent and not directly observable, e.g., the liquidity in a dark or grey venue, or the waiting queues of limit orders. 

In addition, it is also less obvious why the two random components $C$ and $Z$ should be \textit{independent}.  Here perhaps one needs a bit bold revolution that is parallel to the ``\href{\wwwRelativityTheory}{Theory of Relativity}," \textit{in spirit}. For a sizable trading path ripping through a given market (e.g., with an average market participation rate of 30\%, say), why should one still believe in the existence of an ``absolute" market where the rest participants still trade according to a pre-designed and undisturbed Brownian motion? 

There is still some long way to go before reaching a more coherent and matured theory of TCM and more rigorous computational implementations. Notice that the \href{\wwwRSquared}{R-Squared scores} are universally low for  TCM models of the current generation, as low as in the teens or with single digits. TCM 2.0, which is yet to come, can perhaps substantially benefit from modern data analytics as well as machine learning techniques. A \textit{coherent theory} clearly holds the key. On the other hand, for many less liquid FICC securities or ones whose markets are still at their infancy stages, it will take some extra miles to materialize even TCM 1.0.

\subsection{Automated Controls of Automated Trading}

\newcommand{\wwwRegMkt}{https://en.wikipedia.org/wiki/Multilateral_trading_facility}
\newcommand{\wwwMiFID}{https://en.wikipedia.org/wiki/Directive_2014/65/EU}
\newcommand{\wwwKillSwitch}{https://en.wikipedia.org/wiki/Kill_switch}
\newcommand{\wwwOMS}{https://en.wikipedia.org/wiki/Order_management_system}
\newcommand{\wwwEMS}{https://en.wikipedia.org/wiki/Execution_management_system}

Algos differ from many conventional financial products or services. First and foremost, Algos directly access the \href{\wwwNMS}{National Market Systems} (US), \href{\wwwRegMkt}{Regulated Markets} (\href{\wwwMiFID}{MiFID}, Europe), or general national exchanges. Any serious system glitches, operational incidents, or design flaws could generate broad impacts on the regional security prices, major index levels, associated derivative markets, or even the net asset values (NAV) of pension or retirement funds. 

Algos expose their owners, agents, or clients to all major types of risks, be them investment banks, broker-dealers, trading firms or various funds.  The main risk types include, for example, 
\begin{enumerate}[label=(\alph*)]
\item \textbf{regulatory risk} for infringing rules, laws, or regulations on securities, markets, and trading, 
\item \textbf{financial risks} for suffering substantial principal losses as a result of erroneous trading activities, 
\item \textbf{reputational risks} for violating the core principles of financial integrity, or for offering poorly managed products and services to clients, and
\item \textbf{operational and technological risks} for inadequately testing and monitoring trading systems, networks, or servers and data centers, etc.
\end{enumerate}

Because of the autonomous nature, most behaviors of Algos have to be controlled in an \textit{automated} or \textit{low-touch} way, instead of via manual or high-touch interventions. The latter applies only to ultimate controls such as the Emergency Shutdown Procedure (\textit{a.k.a.} the ``\href{\wwwKillSwitch}{Kill Switch}") when the entire Algo system or exchange connections have to be shut down via manual commands (as in Linux) or clicking on-screen ``panic" buttons. 

Controls throughout the life cycles of orders, e.g., incoming parent orders and  child orders at different stages, have to be automated and embedded within the order or execution management systems (\href{\wwwOMS}{OMS}/\href{\wwwEMS}{EMS}). There should be blocks of control codes or scripts \textit{residing within the OMS/EMS} that can automatically police order activities, e.g., parent order acceptance, child order generation, order splitting and routing, and messaging with external exchanges or venues. 

\textit{No senior management teams or clients can feel truly at ease with black-box Algo systems unless it is confirmed that these systems are largely self-regulatory and that comprehensive controls are automated and algorithmic as well.} 

Controls could be kept simple for a small proprietary trading firm who focuses on only a very limited set of securities using a limited set of Algos. For a large-scale investment bank, broker-dealer, or trading firm, however, it is a daunting task to develop a \textit{rigorous} and \textit{effective} control framework that is transparent and auditable, e.g., by either the internal audit teams or external regulators. These firms trade hundreds or thousands of names across multiple asset classes on each business day, relying on tens or hundreds of Algos.

At the minimum, such a control framework means 
\begin{enumerate}[label=(\roman*)]
\item to establish control governance structures or committees within a given firm, 

\item to develop formal control policies and procedures,

\item to construct and maintain a detailed control \textit{inventories}, including some key pieces such as identified risks, proposed controls, actual implementations within the OMS/EMS or beyond, and unit or regression tests that prove the effectiveness of the implemented controls,

\item to clearly delegate and orchestrate the responsibilities within the \href{\wwwThreeLines}{Three Lines of Defense}, including Algo desks, risk management, and independent internal or external audit teams, and

\item to monitor and document the entire life cycles of the controls, including (a) any onboarding requirements for new Algos and associated controls, (b) change management of existing controls, (c) effectiveness and breaching incidents of the established controls, and (d) periodic reviews of the controls.  
\end{enumerate}
\textit{To establish a matured control framework often requires multiple years of commitment and investment from investment banks or trading firms}!

To better elucidate the above discussion, which is somewhat abstract, let us walk through a relatively ``simple" example. 

Suppose for a given Algo named OGLA, among its 280 proposed controls, there is one specific control with identification number CTL-ID9988 which is to limit an incoming parent order to a \textit{compliance limit} of $\Theta = 128$ MM USD. Any incoming order exceeding this limit will be rejected and returned to the trader or client who has submitted it. 

This control reads very self-explanatory and almost trivial. But one should \textit{not} be fooled by its illusory simplicity!  From the control-framework point of view, one can and should challenge it from multiple facets.
\begin{itemize}
\item (\textbf{Ownership}) Who defines this limit of 128 MM? And who are the validators and approvers?
\item (\textbf{Documentation}) What is the rationale in the historical context of this Algo named OGLA? And where is this rationale documented?
\item (\textbf{Data Security}) Within the trading system of Algo OGLA, where is this limit number of ``128 MM USD" stored? And who has the right to access and overwrite it? 
\item (\textbf{Ongoing Monitoring}) In the past quarters, what is the rejection rate under this limit? If the rejection rate has been consistently high, which may have signalled a systematic increment of trading scales from clients (instead of due to fat fingers), should the Algo desk consider to increase this limit for legitimate business? In the same fashion, if the maximum position has been only 30\% of this limit consistently in the past quarters, should the Algo desk consider to lower it in order to more effectively curb fat-finger errors?
\item (\textbf{Exception Handling}) When there is a legitimate reason for a trade to go above this limit, e.g., when both the external client and the internal sales trader(s) have manually communicated about and confirmed such a trade size, what is the emergency procedure for such a trade to legitimately pass the limit check, instead of being rejected outright?
\end{itemize}

The example has been fabricated but the above points are profoundly real. The actual impact could go way beyond the couple of lines that implement such a deceptively trivial control:
\begin{lstlisting}[basicstyle=\small]
  if(order.notional > CTL_ID9988.limit) 
      order.status = REJECTED; ...
\end{lstlisting}

Another major challenge for developing a coherent control framework is that Algos by nature are \textit{dynamic}. In response to emerging trading environments, new client requirements, or novel IT developments, Algo systems are in a constant state of morphing and revamping. It is highly burdensome to scrutinize and approve frequent but legitimate (and occasionally very urgent) changes while maintaining a consistent policy. 

One partial solution is perhaps to replace human approvers and validators by automated algorithms, e.g., via machine learning techniques or artificial intelligence. But this could take away the already shrinking pool of jobs for working daddies and mommies - a ubiquitous wrestle between humans and AIs in the modern era.

\subsection{Full-Scale Testing and Simulation}

\newcommand{\wwwOOP}{https://en.wikipedia.org/wiki/Object-oriented_programming}
\newcommand{\wwwUnitTests}{https://en.wikipedia.org/wiki/Unit_testing}
\newcommand{\wwwRegTests}{https://en.wikipedia.org/wiki/Regression_testing}

\newcommand{\wwwControlFlow}{https://en.wikipedia.org/wiki/Control_flow}
\newcommand{\wwwSwitchFlow}{https://en.wikipedia.org/wiki/Switch_statement}

If analytics and strategies define the mind and soul of an Algo, lines after lines of codes then build up the very flesh and body. Ensuring the healthiness of the body is the  highest priority for Algo development and maintenance. 

The codes embody all the critical functions of Algos, such as messaging with external venues or clients, listening to real-time market trade and quote (TAQ) data, and querying reference data, profiling data, or parameter factories. Most importantly, they implement all the core EMS/OMS logic. The codes as a whole constitute into a complex ecosystem of interacting units.

In general, \href{\wwwOOP}{objective oriented programming} (OOP) offers an effective framework for large-scale code design, components structuring, sharing of common functionalities, and multi-developer collaboration. For example, C++ and Java have been broadly employed as the mainstream languages for Algo development. But sound OOP structuring does not always guarantee bulletproof shelters from coding errors. 

As time passes, any given Algo system has to evolve in order to fix bugs, incorporate novel strategies, or offer new functions and features. Then logic and strategies become more and more involved and coding structures more complex. As a result, an Algo system may become increasingly vulnerable to programming bugs and flaws. 

The \textit{human} factor is a significant source of such potential errors. Developers or strategists come and go. Consequently coding styles may change and many hidden intentions of initial designs (e.g., on classes, variables or functions) may gradually get lost or misused. Formal documentation of all coding details is virtually impossible, while informal in-line commenting is also insufficient to maintain code sanity. 

The other major source of errors arises from all the revamping efforts for expanding new features, functionalities, or products. It is highly nontrivial to ensure an \textit{organic} integration of the new and old codes, especially for large-scale or in-depth projects such as platform migrations or adopting complex quantitative models. Here are two example scenarios when due diligence must be exercised. In reality, one must face all types of challenging scenarios. 
\begin{enumerate}[label=(\alph*)]
\item For instance, inserting a new member function into an existing class which modifies an existing \textit{global} variable could turn very risky without careful examination on how the variable has been utilized in the existing framework. This is especially true when this global variable has been used somewhere else as a control signal for making trade decision such as order splitting or cancellation.

\item New Algo products are often built upon existing modules. These units must be \textit{organically} integrated, instead of being linked perfunctorily. Previously they may have been operating independently. Once being encapsulated under the hood of some new parent logic, these units may have to run in parallel or series. As a result, a responsible developer would have to carefully examine the signals that these modules all listen to, the controls all governed by, as well as the complete flow of cause-effect events. Otherwise, serious glitches could surface under certain trading environments that happen to awaken some previously dormant bugs. 
\end{enumerate}  

\href{\wwwUnitTests}{\textit{Unit Tests}} are designed to verify that individual member functions or task blocks have been coded up as intended. \href{\wwwRegTests}{\textit{Regression Tests}} make sure that these units or other general functionalities remain stable and predictable during rounds of code changes. Regression tests are especially critical for hard compliance controls such as on notional or credit limits. 

These popular and automated tests, if sufficiently comprehensive and accurate, can indeed deliver a high-level of assurance on the soundness of an Algo system.

But an Algo system is not merely an \textit{inorganic} stack of individual units or functionalities. In general, neither unit tests nor regression tests can go \textit{all the way bottom-up} to cover the entire dynamic decision trees embedded within an Algo system. \textit{Most often bugs sneak around in these decision trees where no tests have ever probed.}

As a result, unit and regression tests must be augmented by full-scale simulations of an entire Algo system. This is where the ultimate challenge lies. 

\newcommand{\wwwCircuitBreaker}{https://en.wikipedia.org/wiki/Trading_curb}

First of all, it is highly nontrivial to simulate the dynamics and all possible scenarios of the markets. For example, for testing purposes, one must be able to simulate a sudden trade halt (e.g., as triggered by a \href{\wwwCircuitBreaker}{circuit breaker} rule) and to test/simulate how an Algo system handles the entire life cycle of such a halt. Similarly, for a global market system whose trading hours revolve just as the Earth does, e.g., the FX market, the simulation system must be able to highlight and react to the particular market open and close periods of other regional markets and the companion liquidity spikes. These are merely two examples. 

Even with well-designed mocked market dynamics and event sequencing, the other side of the challenge is to require a simulation system to go through \textit{all} scenario paths that an Algo system can possibly wander through. The dynamic actions of an Algo may involve numerous control or switch statements, e.g., typically coded up by lines like ``\href{\wwwControlFlow}{if-elseif's-else}" or ``\href{\wwwSwitchFlow}{switch}." In addition, they are often cascaded from parent-level requests to children-level responses. The net effect is that a typical Algo amounts to a growing decision tree with numerous branches along the time axis. Failure to simulate through any particular path may expose the Algo to a potentially unregistered bug. But it is a daunting task to ensure that all probably paths be fully visited and simulated. 

Finally, Algo developers should pay extra attention to the testing and simulation of \textit{system capacity}. An Algo system or action that runs smoothly for 5 test names in simulation does not \textit{necessarily} prove that it will behave so in a real trading environment for 500 synchronous names. When this happens, the financial risk could turn very high when an Algo system fails for critical actions like the ``Kill Switch."

\vfill 
\section{Conclusion}

\newcommand{\wwwLowLatency}{https://en.wikipedia.org/wiki/Low_latency_(capital_markets)}

Algorithmic trading and its effective controls have been playing a fundamental role in contemporary financial technologies and services. Optimized trade execution for pension funds, retirements funds and non-profit endowment funds, for example, directly impacts the life of hundreds of millions of main-street citizens. Similarly, automated market making is also vital for maintaining orderly and robust markets by stable liquidity pooling. 

Therefore, it is beneficial for the entire Algo trading community to nurture an open and collaborative culture, as well as to ensure the healthiness and further advancement of ATC.  

This editorial article has been written in this very spirit. We have summarized the current status and challenges facing the nine important facades of ATC. When turned over, the coin of challenges also reveals the other side of exciting opportunities and competing edges in the Algo trading industry. Trading institutions who are committed in making further investments in talents, analytics, technologies, and control frameworks will finally excel. 

Finally, we emphasize that by no means these challenges are claimed to constitute into an \textit{exclusive} list. There exist also some other important ones, for example, with regard to the robustness of data servers and connections, effectiveness of integrating modern data analytics, handling less liquid asset classes or OTC derivatives, and developing ultra \href{\wwwLowLatency}{low-latency} trading systems. 

We also remind readers that the editorial views herein could be inevitably personal and biased.

\section*{Acknowledgments} 

\newcommand{\wwwGSET}{http://www.gset.gs.com/gcl/about/contact.asp}
\newcommand{\wwwJPEquities}{https://www.jpmorgan.com/solutions/cib/markets/global-equities}
\newcommand{\wwwBarcapEquities}{https://www.investmentbank.barclays.com/markets/equities-liquid-markets.html}
\newcommand{\wwwEY}{https://www.ey.com/en_us}
\newcommand{\wwwKPMG}{https://home.kpmg/xx/en/home.html}

The author is very grateful to all the former colleagues at the equities Algo trading teams at both \href{\wwwJPEquities}{J.P. Morgan} and \href{\wwwBarcapEquities}{Barclays}, all the risk and control teams at \href{\wwwGSET}{Goldman Sachs} for electronic trading on all asset classes, as well as all the consultant colleagues from \href{\wwwEY}{E\&Y} and \href{\wwwKPMG}{KPMG} who have played critical roles in disseminating and fusing knowledge and practices on electronic trading. The views in this article are however personal, and by no means imply any endorsement by or representation of these institutions or colleagues. 

\vfill
\hfill
\noindent{\footnotesize\sffamily\color{red!40} Version History: v.202101 - Initial Publication}


\end{document}